\documentclass[12pt]{paper}
\usepackage{amsmath,amssymb,amsfonts,epsfig,graphicx,euscript}%,mathrsfs}

\newcommand{\bse}{\begin{subequations}}
\newcommand{\ese}{\end{subequations}}
\newcommand{\be}{\begin{equation}}
\newcommand{\ee}{\end{equation}}
\newcommand{\bea}{\begin{eqnarray}}
\newcommand{\eea}{\end{eqnarray}}
\newcommand{\ba}{\begin{array}}
\newcommand{\ea}{\end{array}}

\begin{document}
%\hfill%
%\vbox{
%    \halign{#\hfil        \cr
%           IPM/P-2010/014\cr
%                     }
%      }
%\vspace{1cm}
\begin{center}
{ \Large{\textbf{Heavy quarks in the presence of higher derivative
corrections from AdS/CFT}
\\}}
\vspace*{2cm}
%\begin{center}
{\bf K. Bitaghsir Fadafan}\\%
\vspace*{0.4cm}
{\it {Physics Department, Shahrood University of Technology,\\
P.O.Box 3619995161, Shahrood, Iran }}\\
{E-mails: {\tt bitaghsir@shahroodut.ac.ir}}%
\vspace*{1.5cm}
%\end{center}
\end{center}

\vspace{.5cm}
\bigskip
\begin{center}
\textbf{Abstract}
\end{center}
We use the gauge-string duality to study heavy quarks in the
presence of higher derivative corrections. These corrections
correspond to the finite coupling corrections on the properties of
heavy quarks in a hot plasma. In particular, we study the effects of
these corrections on the energy loss and the dissociation length of
a quark-antiquark pair. We show that the calculated energy loss of
heavy quarks through the plasma increases. We also find in general
that the dissociation length becomes shorter with the increase of
coupling parameters of higher curvature terms.

\newpage
\tableofcontents

\section{Introduction}
The experiments of Relativistic Heavy Ion Collisions (RHIC) have
produced a strongly-coupled quark$-$gluon plasma
(QGP)\cite{Shuryak:2004cy}. There are no known quantitative methods
to study strong coupling phenomena in QCD which are not visible in
perturbation theory (except by lattice simulation). A new method for
studying different aspects of QGP is the $AdS/CFT$ correspondence
\cite{Maldacena:1997re,Gubser:1998bc,Witten:1998qj, Witten:1998zw}.
This method has yielded many important insights into the dynamics of
strongly-coupled gauge theories. It has been used to investigate
hydrodynamical transport quantities in various interesting
strongly-coupled gauge theories where perturbation theory is not
applicable \cite{CasalderreySolana:2011us}. Methods based on
$AdS/CFT$ relate gravity in $AdS_5$ space to the conformal field
theory on the four-dimensional boundary. It was shown that an $AdS$
space with a black brane is dual to a conformal field theory at
finite temperature.

The universality of the ratio of shear viscosity $\eta$ to entropy
density $s$
\cite{Policastro:2001yc,Kovtun:2003wp,Buchel:2003tz,Kovtun:2004de}
for all gauge theories with Einstein gravity dual raised the
tantalizing prospect of a connection between string theory and RHIC.
The results were obtained for a class of gauge theories whose
holographic duals are dictated by classical Einstein gravity.
Recently, $\frac{\eta}{s}$ has been studied for a class of CFTs in
flat space with higher derivative corrections
\cite{Brigante:2007nu,Brigante008gz,Kats:2007mq,Neupane:2008dc,Ge:2008ni,Buchel:2008wy}.
In these studies, the effects of  $R^2$ corrections to the
gravitational action in AdS space have been computed and it was
shown that the conjecture lower bound on the $\frac{\eta}{s}$ can be
violated. For example, in the Reissner$-$Nordstr\"{o}m$-$AdS black
brane solution in Gauss$-$Bonnet gravity, the $\frac{\eta}{s}$ bound
is violated and the Maxwell charge slightly reduces the deviation
\cite{Ge:2008ni}. Regarding this study and motivated by the vastness
of the string landscape \cite{Douglas:2006es}, we explored the
modification of the jet quenching parameter and drag force on a
moving heavy quark in the strongly-coupled plasma in
\cite{Fadafan:2008uv}.

Recently, a new higher derivative theory of gravity in
five-dimensional spacetime which contains not only the Gauss-Bonnet
term but also a curvature-cubed interaction introduced
\cite{Myers:2010ru,Oliva:2010eb}. This theory is known as
quasi-topological gravity theory which is thought to be dual to the
large $N$ limit of some conformal field theory without
supersymmetry. Unlike Lovelock gravity, this cubic term is not
purely topological. Therefor it would be useful to consider
curvature-cubed terms as the higher derivative corrections and
investigate behavior of the heavy quarks by means of the $AdS/CFT$
correspondence. Holographic investigation of Quasi-topological
gravity have been done in \cite{Hquasi}. Also it was shown that the
lower bound of the ratio of shear viscosity to density entropy can
be violated in this background \cite{Myers:2010jv}.

In this paper we use the $AdS/CFT$ correspondence to study effect of
higher derivative corrections to the properties of the heavy
quarks.\footnote{ In general, we do not know about forms of higher
derivative corrections in string theory, but it is known that due to
the string landscape one expects that generic corrections can
occur.} One should notice that string theory contains higher
derivative corrections from stringy or quantum effects, and such
corrections correspond to $1/\lambda$ and $1/N$ corrections. In the
case of $\mathcal{N}=4$ super Yang$-$Mills theory, the dual
corresponds to the type $\amalg B$ string theory on $AdS_5\times
S^5$ background. The leading order corrections in $1/\lambda$ arise
from stringy corrections to the low energy effective action of type
$\amalg B$ supergravity, $\alpha'^3 R^4$.

Employing numerical methods, we investigate energy loss and
dissociation length of heavy quarks in section 2 and 3,
respectively. One finds that the energy loss of heavy quarks
increases by increasing higher derivative corrections. Also the
dissociation length becomes shorter with the increase of coupling
parameters of higher curvature terms. We summarize the effects of
these corrections in the last section. In the appendix,
we give a brief review of \cite{Myers:2010ru}.%

\section{Energy loss of heavy quark at finite coupling}

In this section, we investigate the finite-coupling corrections to
the energy loss of a moving heavy quark in the Super Yang-Mills
plasma using the AdS/CFT. These corrections are related to the
curvature corrections to the AdS black brane solution.

The effect of curvature-squared corrections to the drag force on a
moving heavy quark in the Super Yang-Mills plasma is investigated in
\cite{Fadafan:2008gb}. It is shown that the corrections to the drag
force depend on the velocity of heavy quark. This dependance is such
that for $v > v_c$ including the corrections increase the drag
force. This means that at the critical velocity $v_c$ the
curvature-squared corrections have the minimum effect on the drag
force. For the particular case of Gauss-Bonnet gravity, we do not
expect a critical velocity \cite{Fadafan:2008gb}. Also in this
background, the drag force is larger than the $\mathcal{N}=4$ case
if $\lambda_{}$ (Gauss-Bonnet gravity constant) is positive while it
is smaller than the $\mathcal{N}=4$ case if $\lambda_{}$ is
negative.

Now we continue with considering curvature-cubic corrections. Our
purpose is finding a general rule for considering higher derivative
terms. We use the proposal of \cite{Myers:2010ru,Oliva:2010eb} and
study new higher derivative theory of gravity in five-dimensional
spacetime which contains not only the Gauss-Bonnet term but also a
curvature-cubed interaction.

We should emphasize that in the case of these corrections, one can
not predict a result for $\mathcal{N}=4$ SYM because the first
higher derivative correction in weakly curved type IIB backgrounds
enters at order ${\cal{R}}^4$. These corrections on the drag force
have been studied in \cite{VazquezPoritz} and it was found that the
drag force for a heavy quark moving through $\mathcal{N}=4$ SYM
plasma is generally enhanced by the leading correction due to finite
't Hooft coupling. We will compare our results with this observation
and interestingly find a general rule for curvature corrections.

\subsection{Set up of calculations}
In the framework of $AdS/CFT$, an external quark is represented as a
string dangling from the boundary of $AdS_5-$Schwarzschild and a
dynamical quark is represented as a string ending on flavor D7-brane
and extending down to some finite radius in $AdS$ black brane
background. We consider the $AdS$ black hole solution in
quasi-topological gravity \footnote{one finds a quick review of this
background in the appendix.} as

\be %
ds^2=r^2\left(-N^2\,f(r)dt^2+d\vec{x}^2\right)+\frac{dr^2}{r^2\,f(r)},\label{metric1}
\ee%
notice that we work in units where the radius of $AdS$ is one. Here
$r$ denotes the radial coordinate of the black brane geometry and
$t, \vec{x}$ label the directions along the boundary at the spatial
infinity. In these coordinates the event horizon is located at $r_h$
and it is found by solving $f(r_h)=0$ equation. The boundary is
located at infinity and the geometry will be as asymptotically
$AdS$. The constant $N^2$ specifies the speed of light of the
boundary gauge theory and one can choose it to be unity. We
name $f(r)$ at the boundary where $r\rightarrow\infty$, as $f_{\infty}$ and one finds that%
\be N^2=\frac{1}{f_\infty},\label{N}\ee%
The temperature of the hot plasma is given by the Hawking
temperature of the black hole
\begin{equation}
T=\frac{N\,r_h}{\pi}\label{T}.
\end{equation}

The relevant string dynamics is captured by the Nambu-Goto action
\be S=-\frac{1}{2 \pi \alpha'}\int d\tau d\sigma\sqrt{-det\,g_{ab }
},\label{Numbo-Goto} \ee%
where the coordinates $(\sigma, \tau)$ parameterize the induced
metric $g_{ab}$ on the string world-sheet and $X^\mu(\sigma, \tau)$
is a map from the string world-sheet into the space-time. Defining
$\dot X = \partial_\tau X$, $X' =
\partial_\sigma X$, and $V \cdot W = V^\mu W^\nu G_{\mu\nu}$ where
$G_{\mu\nu}$ is the AdS black hole solution in Quasi-topological
gravity \eqref{metric1}, then
\begin{equation}
-det\,g_{ab }=(\dot X \cdot X')^2 - (X')^2(\dot X)^2.
\end{equation}
We follow \cite{Herzog:2006gh,Gubser:2006bz} and focus on the dual
configuration of the external quark moving in the $x$ direction on
the plasma. The string in this case, trails behind its boundary
endpoint as it moves at constant speed $v$ in the $x$ direction
\begin{equation}
x(r,t)=v t+\xi(r),\,\,\,\,\,  y=0,\,\,\, z=0.
\end{equation}
One finds the lagrangian in the static gauge $(\sigma=r, \tau=t)$ as
follows
\begin{equation}
\mathcal{L}=\sqrt{-det\,g_{ab}}=N^2+r^4 N^2 f(r)
x'^2-\frac{\dot{x}^2}{f(r)}, \label{lag}
\end{equation}
The equation of motion for $\xi$  implies that $\frac{\partial
L}{\partial \xi'}$ is a constant. We name this constant as
$\Pi_{\xi}$ and solve this relation for $\xi'$, the result is
\begin{equation}
\xi'^2=\frac{\left(\frac{\Pi_{\xi}^2}{f(r)}\right)\left( - N^2
f(r)+v^2 \right)}{r^4N^2f(r)\left(-r^4 N^2 f(r)+\Pi_{\xi}^2
\right)}.\label{Xi}
\end{equation}
We are interested in a string that stretches from the boundary to
the horizon. In such a string, $\xi'^2$ remains positive everywhere
on the string. Hence both numerator and denominator change sign at
the same point and with this condition, one finds the constant of
motion $\Pi_{\xi}$ in terms of the critical value of $r_c$ as follows%
\be \Pi_{\xi}=v\,r_c^2, \label{pi}\ee%
The drag force that is experienced by the heavy quark is calculated
by the current density for momentum along $x^1$ direction. After
straightforward calculations, the drag force is easily simplified in
terms of $\Pi_{\xi}$
\begin{equation}
F=-\frac{1}{2 \pi \alpha'} \Pi_\xi\label{generaldrag}.
\end{equation}
As a result, to find the drag force one should find the constant of motion, $\Pi_{\xi}$ from \eqref{pi}.
Numerator and denominator in (\ref{Xi}) change sign at $r_c$ and it can be
 found by solving this equation%
\be f(r_c)-\frac{v^2}{N^2}=0. \label{dragroot}\ee%

As it is clear in the appendix, Gauss-Bonnet coupling and
curvature-cubed interaction constant are $\lambda$ and $\mu$,
respectively and the precise form of $f(r)$ depends on $\lambda$ and
$\mu$. It was found that there are three different $AdS$ black hole
solutions in quasi-topological gravity which are determined by
$f_1(r), f_2(r)$ and $f_3(r)$ in \eqref{f}. Then for different
values of coupling constant $\lambda$ and $\mu$, one should choose
appropriate form of $f(r)$ from \eqref{f} and solve
\eqref{dragroot}. However \eqref{dragroot} is complicated one can
solve it numerically. Then,
 we assume different values for $\mu$ and $\lambda$ and discuss behavior of the drag force
in terms of these coupling constants.

\subsection{Positive couplings}
We assume both coupling parameters $\mu$ and $\lambda$ are positive.
As pointed out in \cite{Myers:2010ru}, for this case, only $f_3(r)$
in \eqref{f} leads to a stable AdS black hole solution. The drag
force versus the velocity of the heavy quark  has been plotted in
Fig. 1. In the right and left plots of this figure, Gauss-Bonnet
coupling constant is $\lambda=0.01$ and $\lambda=0.20$,
respectively. Also different values of cubic-curvature coupling
interaction are assumed.

As one finds from \cite{VazquezPoritz}, by increasing $\lambda$ the
value of the drag force increases. This behavior of the drag force
is clearly seen in these plots. One finds that by increasing
Gauss-Bonnet coupling constant from $\lambda=0.01$ to
$\lambda=0.20$, the drag force increases. In the plots of Fig. 1,
one finds that by increasing $\mu$ the value of the drag force also
increases. Though at the small velocities, the cubic-curvature
interactions have minimum effect on the drag force. As a result, the
main effect of increasing cubic-curvature coupling constant is
increasing the drag force value. This is the same as the case of
$R^2$ and $R^4$ case \cite{Fadafan:2008gb,VazquezPoritz}. We should
check this result in the case of non-positive $\mu$ and $\lambda$.
%%%%%%%%%%%%%%%%%%%%%%%%
\begin{figure}
  \centerline{\includegraphics[width=3in]{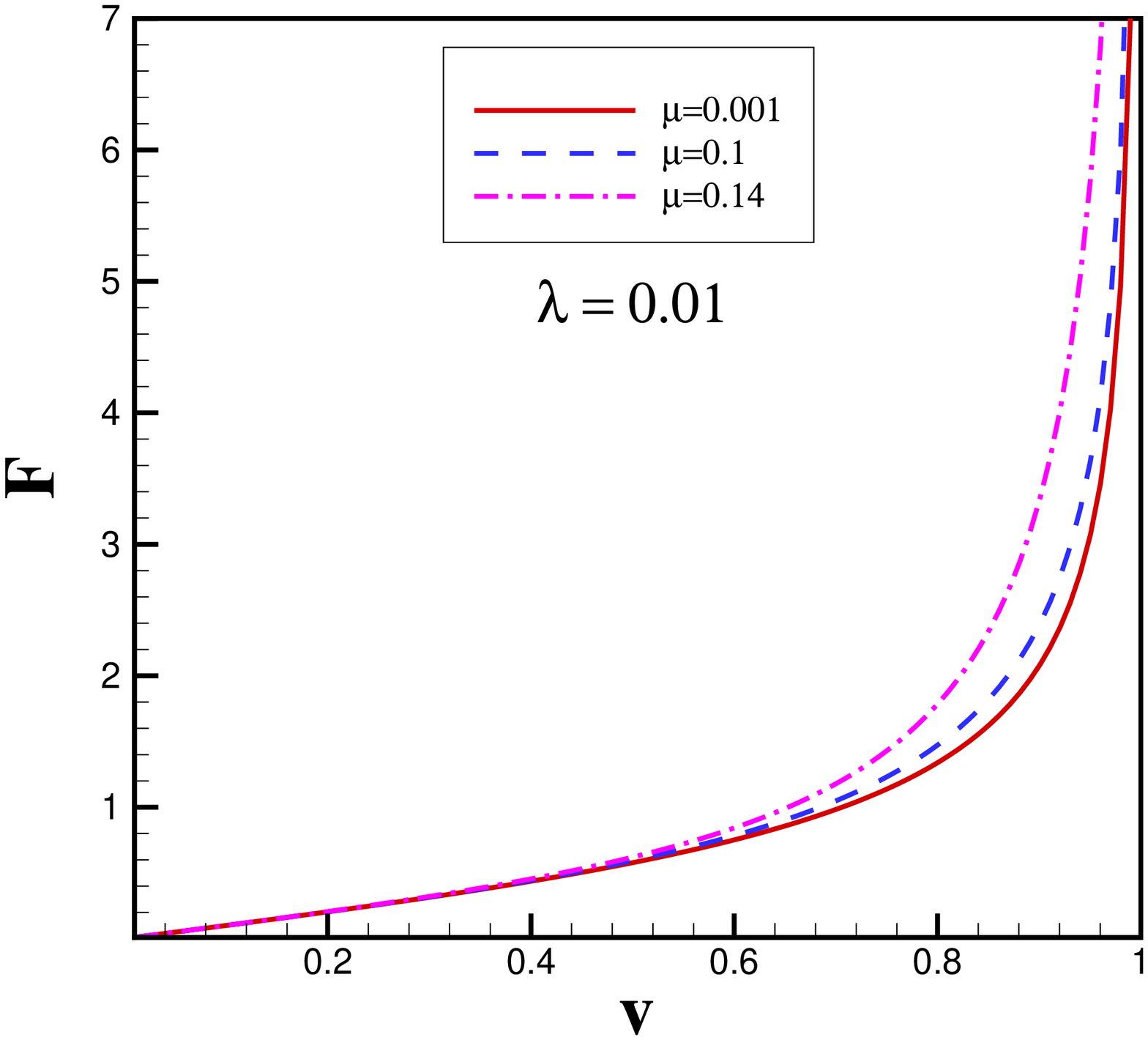},\includegraphics[width=3in]{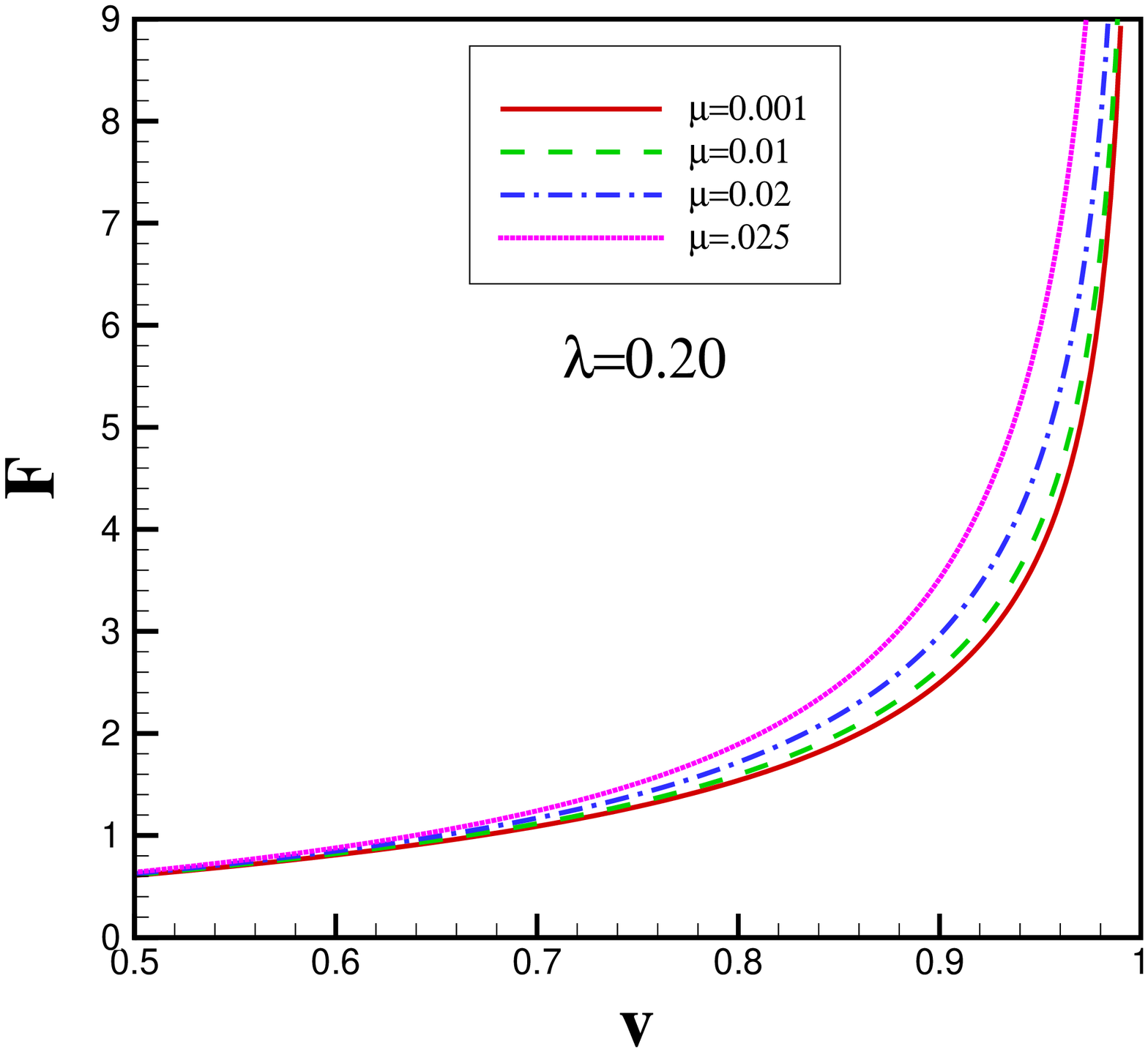}}
  \caption{The drag force versus the velocity of the heavy quark
for \emph{positive} values of cubic-curvature coupling $\mu$ at
fixed \emph{positive} Gauss-Bonnet coupling constant.}\label{fig}
\end{figure}
%%%%%%%%%%%%%%%%%%%%%%%%%
\subsection{non-positive couplings}
Now we intend to study the effect of the non-positive coupling
constants $(\lambda,\mu)$ to the drag force. Three distinct $AdS$
black hole backgrounds are discussed in \eqref{f}. These solutions
for different regimes of the parameter space of $(\lambda,\mu)$ are
discussed in the table 1 of \cite{Myers:2010ru}. As it is explained
in this table, to study the non-positive coupling constants, one
needs $f_1(r),f_2(r)$ and $f_3(r)$ from \eqref{f}. In the case of
positive $\mu$ and negative $\lambda$, only $f_3(r)$ in \eqref{f}
leads to a stable AdS black hole solution. In Fig. 2, we assume
$\lambda=-0.2$ and $\mu=+0.01,+0.02,+0.25$ and plot the drag force
versus the velocity of the heavy quark. Also here, one finds that by
increasing $\mu$ the value of drag force increases. One should
notice that at the small velocities, the cubic-curvature corrections
have the minimum effects. Therefor we confirm the previous result.
If one assumes negative $\mu$ and positive $\lambda$, also finds
that for larger $\mu$ the value of the drag force becomes larger.

\subsection{analytic solution}
Fortunately, we find an analytic result for the drag force in the
special case of
$\mu=-\frac{\lambda^2}{3}$ which corresponds to $p=0$ in \eqref{p}, as%
\be F_{R^2+R^3}=-\frac{1}{2\pi\alpha'}\left(
\frac{\sqrt{3}\,\pi^2\,T^2\,v}{N^{\frac{1}{2}}\sqrt{-v^6\,\lambda^2+3
v^4\,
\lambda\, N-3\,v^2\, N^2+3 N^3}} \right),\label{dragR3} \ee%
where $N$ is defined in \eqref{N}.

It would be interesting to compare the drag force in the presence of
higher derivative corrections with the case of $\mathcal{N}=4$
strongly-coupled SYM plasma $F_{\mathcal{N}=4}$. The authors of
\cite{Gubser:2006bz,Herzog:2006gh} have obtained
\begin{equation}
F_{\mathcal{N}=4}=-\left(\frac{\pi\,\sqrt{\tilde{\lambda}}\,T_0^2}{2}\right)\,\frac{v}{\sqrt{1-v^2}}.\label{dragN4}
\end{equation}
where $\tilde{\lambda}$ is 't Hooft coupling\footnote{Notice that
$\alpha'^{-2}=\tilde{\lambda}$.} and $T_0$ is the temperature of AdS
black hole solution without any corrections. Let us consider the
case of $\lambda\rightarrow0$ in \eqref{dragR3}. In this limit, one
does not consider any correction in the action (\ref{I}) and finds
that the drag force is nothing but the drag force in the case of
$\mathcal{N}=4$ strongly-coupled SYM plasma $F_{\mathcal{N}=4}$.

The analytical result in \eqref{dragR3}, shows the effect of the
higher derivative corrections on the drag force. It is clearly seen
that the corrections appear in the denominator of \eqref{dragR3} and
as a result the drag force increases.
%%%%%%%%%%%%%%%%%%%%%%%%
\begin{figure}
  \centerline{\includegraphics[width=3in]{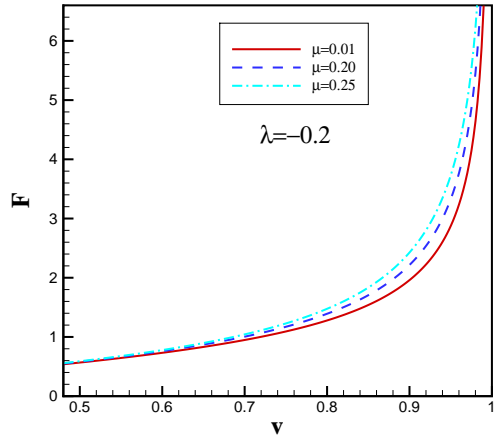}}
  \caption{The drag force versus the velocity of the heavy quark
for \emph{positive} values of cubic-curvature coupling $\mu$ at
fixed \emph{negative} Gauss-Bonnet coupling constant $\lambda
$.}\label{fig}
\end{figure}
%%%%%%%%%%%%%%%%%%%%%%%%%

\section{dissociation length of quark-antiquark pair at finite coupling
}

In this section we investigate the effect of the higher derivative
terms to the dissociation length of quark-antiquark pair.

In the usual fashion, the two endpoints of the classical open string
at the boundary are seen as a quark and antiquark pair which may be
considered as a meson \cite{Erdmenger:2007cm}. Based on lattice
results and experiments, it is found that the meson shows
interesting behavior as the temperature of the plasma increases. It
is known that heavy quark bound states can survive in a QGP to
temperatures higher than the confinement/deconfinement transition
\cite{de Forcrand:2000jx}. Thermal properties of \emph{static}
quark-antiquark systems have been studied in \cite{Maldacena:1998im,
Rey} in an AdS-Schwarzschild black hole setting using the AdS/CFT
correspondence. In \cite{AliAkbari:2009pf}, a rotating
quark-antiquark in the presence of higher derivative corrections is
studied. In the case of Gauss-Bonnet corrections, it is shown that
as the Gauss-Bonnet coupling constant $\lambda_{}$ increases the
string endpoints become less separated {\em{i.e.}} the radius of the
rotating open string at the boundary decreases but the tip of the
U-shaped string does not change considerably.%

The heavy quark potential in the presence of curvature-squared
corrections is calculated in \cite{Noronha:2009ia}. It is shown that
the potential can be calculated as a power series in $LT<<1$, where
$T$ is the temperature of the hot plasma. One finds that at fixed
temperature, as the Gauss-Bonnet coupling constant $\lambda_{}$
increases the interquark distance $L$ decreases. It would be
interesting to investigate this observation in the case of higher
derivative corrections. To do this, we consider Quasi-topological
gravity and study effect of curvature-cubed corrections to the
dissociation length. Because of the complicated feature in this
background, we use numerical methods.

\subsection{dissociation length from AdS/CFT}

To find the dissociation length, one should study the heavy quark
potential, $V_{q\bar{q}}(L)$,\footnote{We call quark-antiquark
potential in \eqref{potential} as "heavy quark potential".} where
$L$ is the distance between two quarks \cite{Antipin}. One finds
that the heavy quark potential can be \emph{negative},
\emph{positive} or \emph{zero}. If $V_{q\bar{q}}(L) <0 $, the
dominant string configuration becomes the one for the U-shaped
string which can be interpreted as a heavy meson. When
$V_{q\bar{q}}(L)>0 $, the heavy meson dissociates to two free quarks
and the string configuration changes. This phenomena happens at
special length $L=d$ which is obtained from $ V_{q\bar{q}}(L=d)=0 $.
We call $d$ as a dissociation length. Thus by studying the heavy
quark potential in the quasi-topological gravity, we will find the
effect of higher derivative terms to this quantity.

The heavy quark potential is given by the expectation value of the
following static Wilson loop%
\be W(C)=\frac{1}{N}Tr\,P\,e^{i\,\int A_\mu dx^\mu},\ee%
where $C$ denotes a closed loop in spacetime and the trace is over
the fundamental representation of $SU(N)$ group. We consider a
rectangular loop along the time coordinate $t$ and spatial extension
$L$. The static heavy quark potential is related to the expectation
value of this rectangular Wilson loop in
the limit of $t\rightarrow\infty$,%
\be \langle W(C)\rangle \sim e^{-t\,V_{q\bar{q}}(L)} ,\label{potential}\ee%

This expectation value can be calculated from $AdS/CFT$
correspondence \cite{Maldacena:1998im,Rey}. In this set up, one
should consider an infinitely massive quark in the fundamental
representation of $SU(N)$ group in $\mathcal{N}=4$ Yang-Mills gauge
theory. This quark is dual to a classical string hanging down to the
horizon from a probe brane at the boundary. The classical string
hanging in the bulk space and connecting two endpoints has a
characteristic U-shaped. We name $r_*$ as the tip of the U-shaped
string and we let it to define the nearest point between the string
and the horizon of the black hole; i.e. $r_*>r_h$. Let us emphasize
that for non-physical states we would have $r_*<r_h$
\cite{Maldacena:1998im}.

The dynamics of the U-shaped string is given by the Euclidean
version of the Nambu-Goto action in \eqref{Numbo-Goto}. To calculate
the heavy quark potential, one has to subtract the infinite
self-energy of two independent heavy quarks and from the $AdS/CFT$
correspondence. These massive quarks are dual to two straight
strings that extend from the probe brane at the boundary to the
horizon. The regularized action is shown by $\bigtriangleup S$ and it is related to the expectation value of Wilson loop in \eqref{potential} by this equation %
\be \langle W(C)\rangle \sim e^{-\bigtriangleup S} ,\ee%
As a result, the heavy quark potential is%
\be V_{q\bar{q}}(L)=\frac{\bigtriangleup S}{t}. \ee%

The heavy quark potential in the vacuum and in the strongly coupled
$N=4$ SYM gauge theory was found in \cite{Maldacena:1998im}%
\be V_{q\bar{q}(L)}=-\frac{4 \pi^2 \sqrt{\lambda}}{\Gamma(1/4)^2}\left(\frac{1}{L}\right).\ee%

We consider $X^\mu=(t,x,0,0,r(x))$ for the coordinates of U-shaped
string in the static gauge $\sigma=x,\, \tau=t$. As a result, the
Euclidean version of Nambu-Goto action in \eqref{Numbo-Goto} can be
found as
\be S=\frac{N\,t}{2\pi \alpha'} \int \, dx \sqrt{r^4\, f(r)+r'^2},
\ee
Notice that $r$ depends on $x$. The Hamiltonian density of this
action is constant and it is
\be H=-\frac{N\,t}{2\pi \alpha'}\frac{r^4\,f(r)}{\sqrt{r^4\,
f(r)+r'^2}}, \ee
This constant is found at special point $r(0)=r_*$, where $r'_*=0$,
as
\be H=-\frac{N\,t}{2\pi \alpha'}\sqrt{r_*^4\, f(r_*)}. \ee
Then it is possible to find $L$ as follows
\be \frac{L}{2}=\int_{r_*}^\infty\,dr\left(\frac{1}
{r^4\,f(r)\left(\frac{r^4\,f(r)}{r_*^4\,f(r_*)}-1\right)}\right)^{1/2}.\label{L}
\ee
Finally, the heavy quark potential is given by%
\be
V_{q\bar{q}}(L)=\frac{\,N}{\pi\alpha'}\,\int_{r_*}^\infty\,dr\left(
\left(\frac{\frac{r^4\,f(r)}{r_*^4f(r_*)}}
{\frac{r^4\,f(r)}{r_*^4f(r_*)}-1}\right)^{\frac{1}{2}}-1\right)-
\frac{\,N}{\pi\alpha'}\,\int_{r_h}^{r_*}\,dr.\label{potential2}\ee

We intend to study the effect of the higher derivative corrections
to the heavy quark potential in \eqref{potential2} and the
interquark distance in \eqref{L}. For different values of coupling
constants $(\lambda,\mu)$, one should consider three distinct $AdS$
black hole backgrounds which are discussed in \eqref{f}. However, we
can not solve \eqref{potential2} and \eqref{L} analytically and we
have to resort to numerical methods. Also the coefficient
$\frac{N\,}{\pi\alpha'}$ does not play any role in our physical
discussion.

\subsection{Numerical Solutions}

We illustrate behavior of $V_{q\bar{q}}(L)$ as a function of $L$ at
fixed temperature $(r_h=1)$ in Fig. 3. It is clearly seen that there
is a maximal interquark distance, $L_{max}$. It has been shown that
for $L<L_{max}$ there are two kinds of strings; long strings and
short strings
\cite{Friess:2006rk,Avramis:2006nv,Avramis:2007mv,Sfetsos:2008yr}.
These strings correspond to the upper and lower parts of
$V_{q\bar{q}}(L)$ in Fig. 3, respectively. The stability analysis
has shown that short strings are favorable
\cite{Friess:2006rk,Avramis:2006nv,Avramis:2007mv,Sfetsos:2008yr}.
One concludes that only the lower part is physical\cite{Rey}.%

By analyzing Fig. 3., we investigate behavior of the dissociation
length for different values for $\mu$ and $\lambda$. In this figure,
the heavy quark potential \eqref{potential2} is plotted versus the
interquark distance \eqref{L}. We take that different values of
cubic-curvature coupling $\mu$ while the Gauss-Bonnet coupling
constant $\lambda $ is fixed in each frame. Notice that in this case
the corresponding black hole backgrounds are specified by $f_3$. In
this figure, from left to right the Gauss-Bonnet coupling constant
$\lambda$ is increasing, $\lambda=-0.2, 0.01$ and $0.2$. By
increasing Gauss-Bonnet coupling constant, the dissociation length
of meson decreases. This phenomena has been found also in the case
of a rotating meson \cite{AliAkbari:2009pf}.
%%%%%%%%%%%%%%%%%%%%%%%%
\begin{figure}
  \centerline{\includegraphics[width=2.2in]{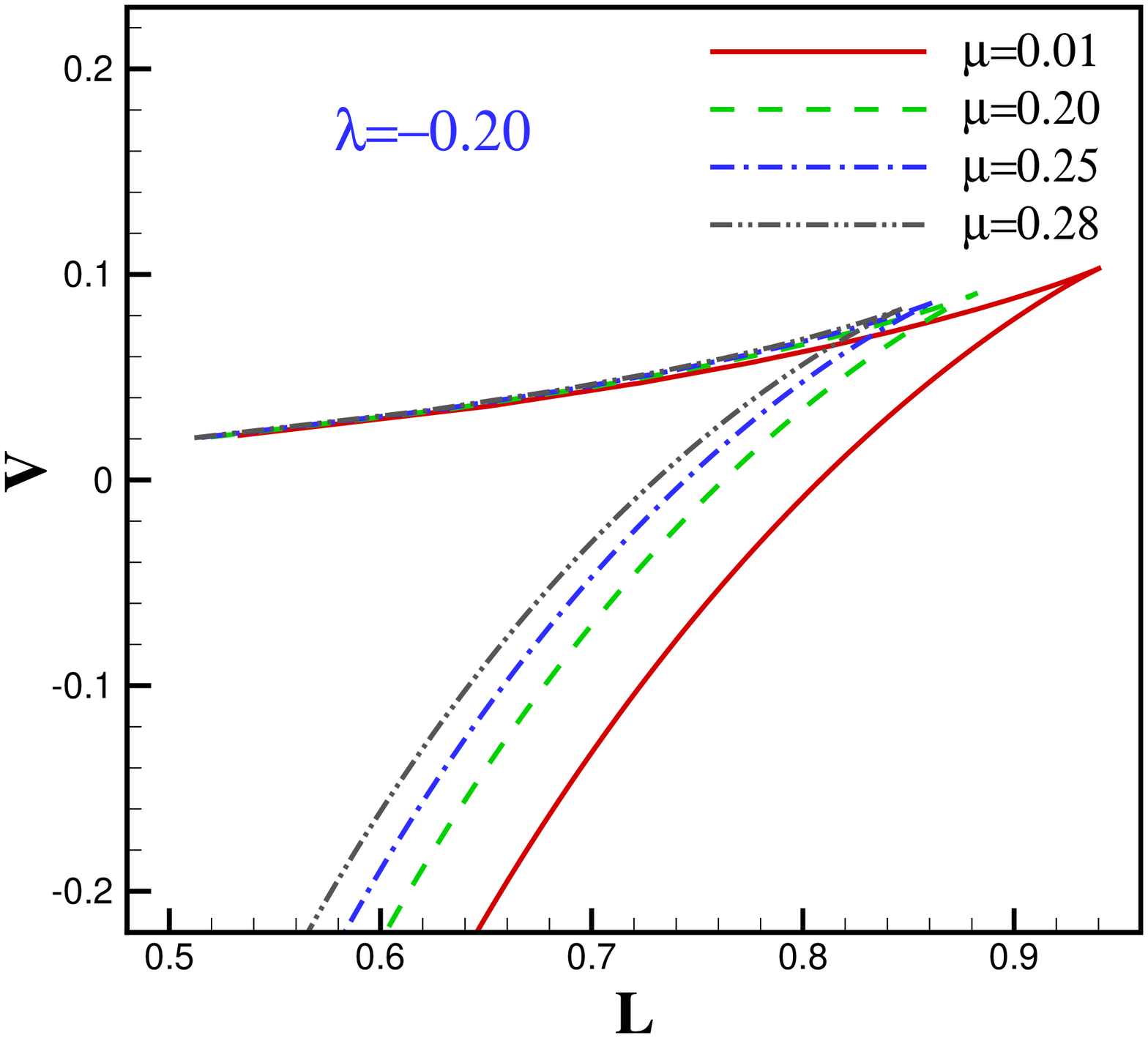}\includegraphics[width=2.2in]{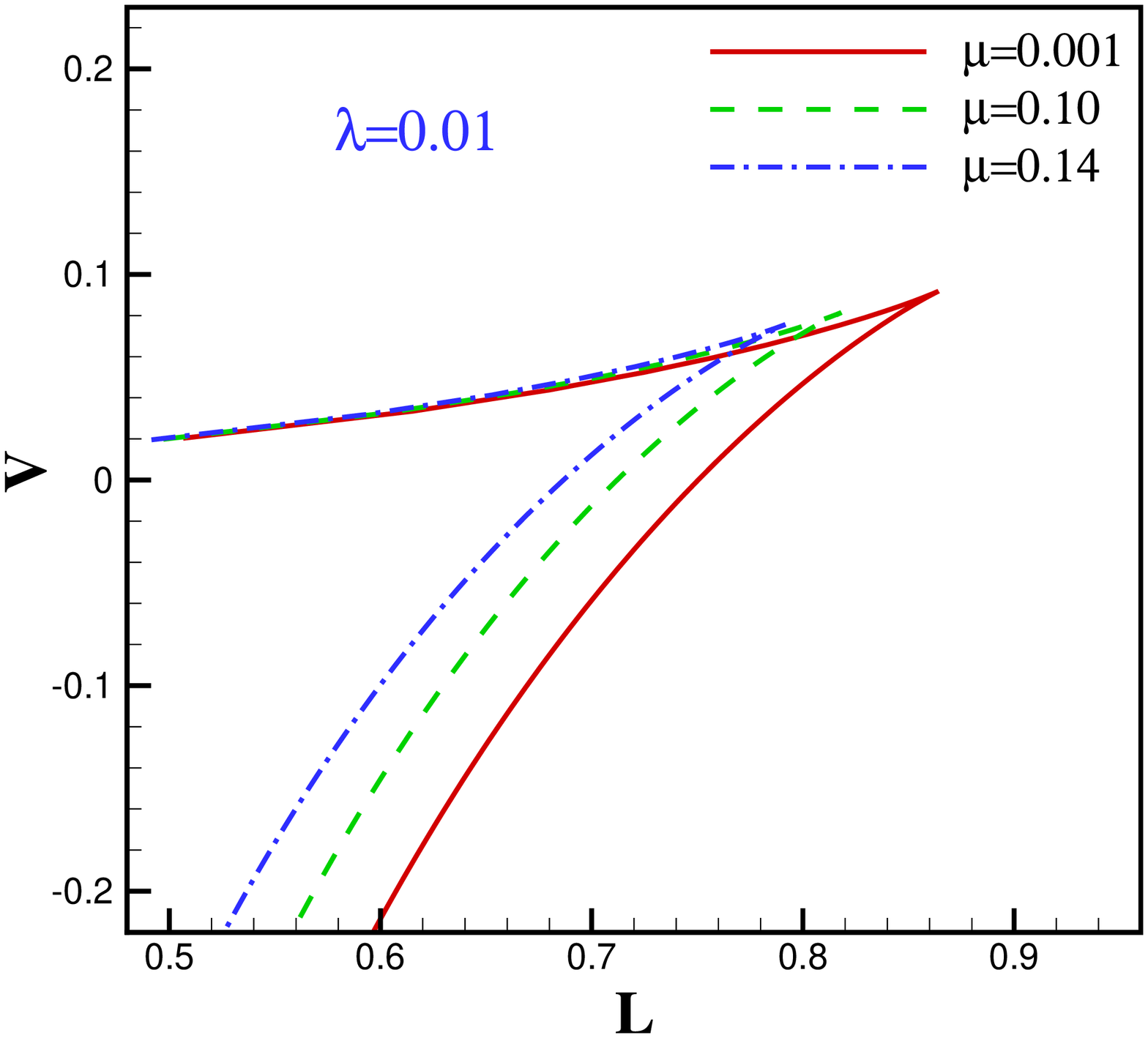}\includegraphics[width=2.2in]{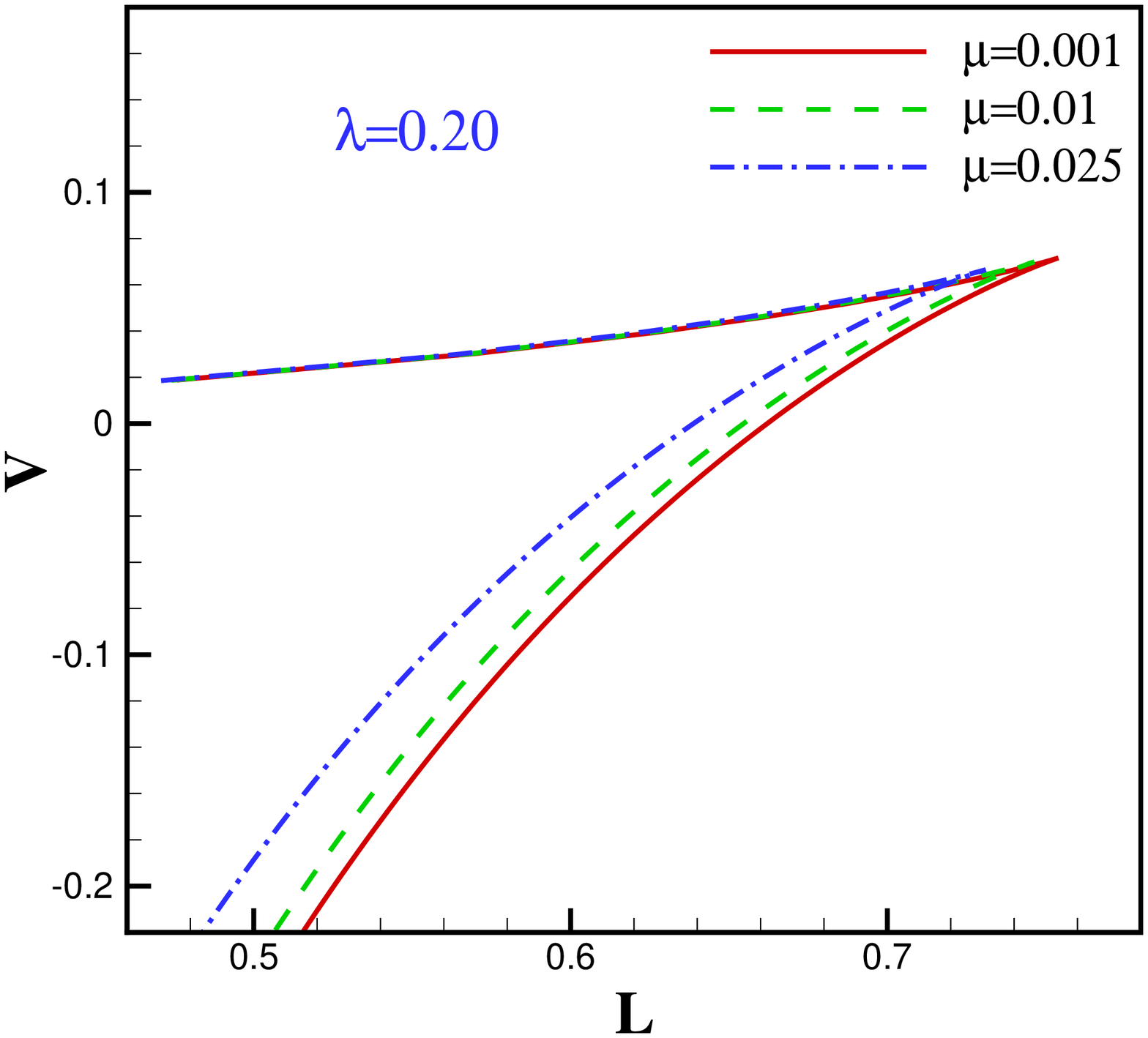}}
  \caption{The heavy quark potential versus the interquark distance
for different values of cubic-curvature coupling $\mu$ at fixed
Gauss-Bonnet coupling constant $\lambda $. Left:$\lambda=-0.2$.
Middle:$\lambda=0.01$. Right: $\lambda=0.2$.}\label{fig}
\end{figure}
%%%%%%%%%%%%%%%%%%%%%%%%%

What is the effect of increasing cubic-curvature coupling $\mu$
while Gauss-Bonnet coupling $\lambda$ is fixed? In each plot of Fig.
3, $\lambda$ is fixed and $\mu$ is increasing. For example in the
left plot of this figure $\lambda=-0.20$ and $\mu=0.01, 0.20, 0.25$
and $0.28$. One can see that the interquark distance decreases by
increasing $\mu$. This observation is clearly seen in the middle and
right plots of Fig. 3, too. Therefor by increasing cubic-curvature
coupling, the dissociation length of meson decreases.

As we pointed out, there are three distinct AdS black hole
backgrounds which correspond to $f_1(r), f_2(r)$ and $f_3(r)$ in
\eqref{f}. In the case of $\lambda<0$ and $\mu<0$ one should
consider $f_1(r)$. We show the heavy quark potential versus the
interquark distance in the left plot of Fig. 4. In this plot,
$\lambda=-0.9$ and from right to left curve $\mu$ is increasing from
$-0.2, -0.1$ to $-0.01$. As before, one finds that the dissociation
length decreases by increasing the cubic-curvature constant $\mu$.
One should notice that the rate of decreasing is not so large. In
the case of $\lambda>0$ and $\mu<0$, one should choose $f_2(r)$ to
investigate behavior of the heavy quark potential versus the
interquark distance. We show the result in the right plot of Fig. 4.
In this plot $\lambda=0.20$ and $\mu$ is increasing from $-0.01$ to
$-0.0001$. It is clearly seen that by increasing $\mu$, the
dissociation length decreases. This observation is consistent with
what we see in Fig. 3.
%%%%%%%%%%%%%%%%%%%%%%%%
\begin{figure}
  \centerline{\includegraphics[width=3.in]{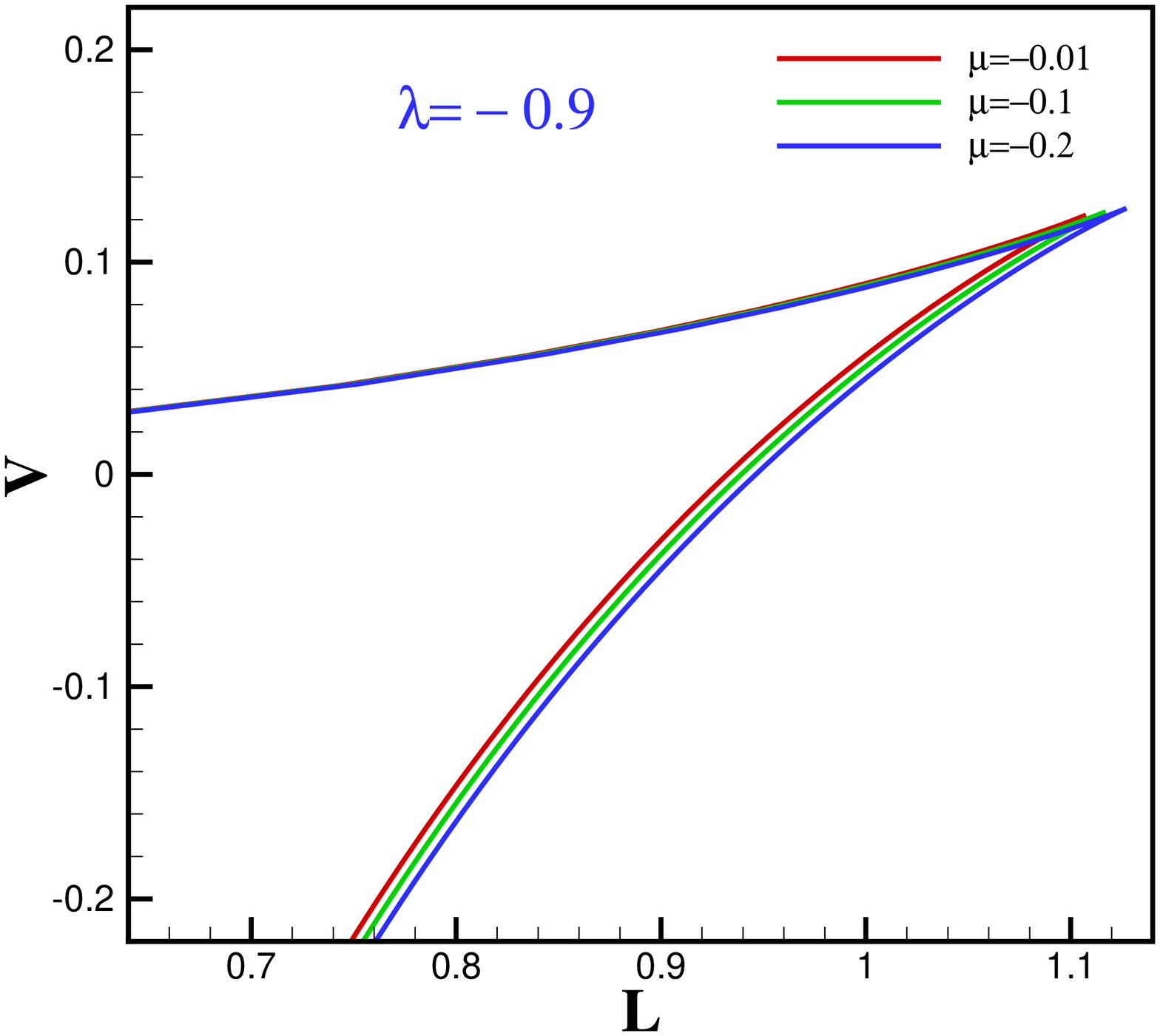}\includegraphics[width=3.in]{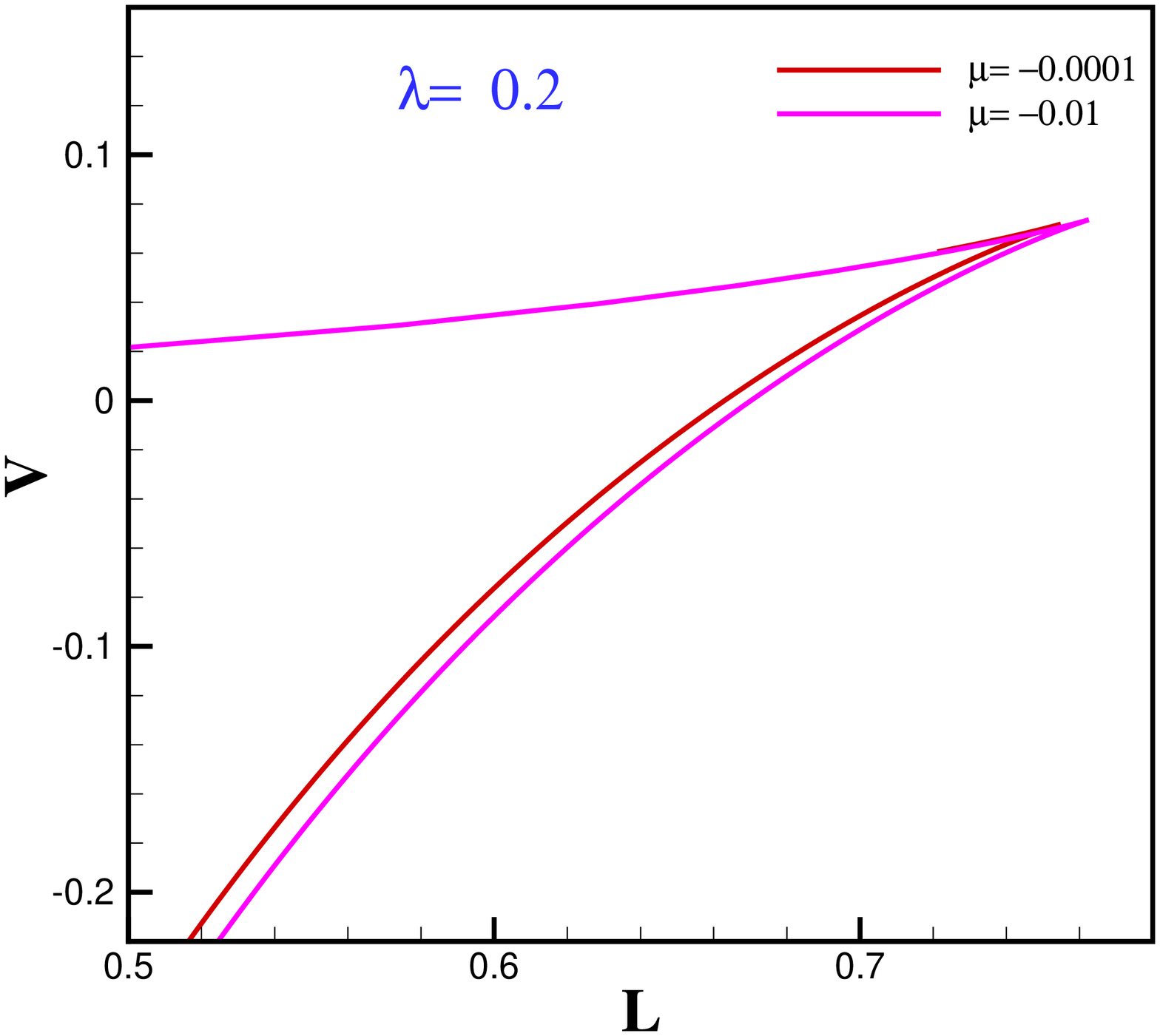}}
  \caption{The heavy quark potential versus the interquark distance.
  Left:$\lambda=-0.9$ and from right to left
  $\mu=-0.2,\,-0.1,-0.01$. Right: $\lambda=0.2$ and from right to left
  $\mu=-0.01,-0.0001$. }\label{fig}
\end{figure}
%%%%%%%%%%%%%%%%%%%%%%%%%
One infers that as Gauss-Bonnet coupling constant $\lambda$
increases the interquark distance decreases. Also at fixed $\lambda$
by increasing cubic-curvature constant $\mu$ the interquark distant
 decreases. As a result, including the higher derivative
corrections decrease the dissociation length.

\section{Conclusion}

The higher derivative corrections on the gravity side correspond to
finite coupling corrections on the gauge theory side. The main
motivation to consider these corrections comes from the fact that
string theory contains higher derivative corrections arising from
stringy effects. On the gauge theory side, computations are exactly
valid when the 't Hooft coupling constant goes to infinity
($\tilde{\lambda}=g_{YM}^2N\rightarrow\infty$). An understanding of
how these computations are affected by finite $\lambda$ corrections
may  be essential for more precise theoretical predictions.

Although AdS/CFT correspondence is not directly applicable to QCD,
one expects that results obtained from closely related non-abelian
gauge theories should shed qualitative (or even quantitative)
insights into analogous questions in QCD. This has motivated much
work devoted to study various properties of thermal SYM theories
like the hydrodynamical transport quantities. In this paper we have
studied the energy loss and the interquark-antiquark distance in the
presence of higher derivative terms. We have considered the
cubic-curvature terms which is known as quasi-topological gravity.

We calculated the energy loss of heavy quark and from numerical
analysis, found that the drag force increases. Fortunately, we found
an analytical result in \eqref{dragR3} which confirms our result.

We introduced the heavy quark potential in \eqref{potential}. As it
is seen from Fig. 3, the heavy quark potential is \emph{negative},
\emph{positive} or \emph{zero}. By studying the zero case, we
investigated effect of higher derivative terms in quasi-topological
gravity on the dissociation length. We found that the
interquark-antiquark distance becomes shorter with the increase of
coupling parameters of higher curvature terms. This result is
consistent with the case of rotating quark-antiquark pair in
\cite{AliAkbari:2009pf}. We can therefore conclude that the higher
curvature corrections make the dissociation length shorter.
Interestingly, the subleading term of the strong coupling expansion
of the heavy quark potential in a $\mathcal{N}=4$ SYM plasma is
studied in \cite{Zhang:2011zj}. It is also found that this
correction reduces the magnitude of the heavy quark potential and
leads to a smaller screening radius.

\section*{Acknowledgment}
We would like to thank E. Azimfard for helpful discussions and
specially thank M. Ali-Akbari and M. Sohani for reading the
manuscript and useful comments.
%\appendix{Review of Quasi-topological gravity}
\section{Review of Quasi-topological gravity}

In this appendix we give a brief review of the quasi-topological
gravity in five-dimensional spacetime \cite{Myers:2010ru}. The
bulk action is given by%
\be I=\frac{1}{16 \pi G_5}\int dx^5
\sqrt{-g}\left(R-\Lambda+\frac{\lambda L^2}{2}\chi_4+\frac{7 L^4\mu
}{8} Z_5\right),\label{I}
\ee%
where $\lambda$ and $\mu$ are Gauss-Bonnet coupling and
curvature-cubed interaction constant, respectively. The negative
cosmological constant is related to radius of AdS space by
$\Lambda=-\frac{12}{L^2}$. The
curvature-squared interaction is given by $\chi_4$ as%
\be \chi_4=R^2-4R_{\mu\nu}R^{\mu\nu}+R_{\mu\nu\rho\sigma}R^{\mu\nu\rho\sigma},\ee%
and $Z_5$ is the new curvature-cubed interaction %
\begin{eqnarray} %
Z_5&=&R_{\mu\nu}^{\,\,\,\,\,\,\rho\sigma}R_{\rho\sigma}^{\,\,\,\,\,\,\alpha\beta}R_{\alpha\beta}^{\,\,\,\,\,\,\mu\nu}
+\frac{1}{14}\left( 21
R_{\mu\nu\rho\sigma}R^{\mu\nu\rho\sigma}\,R-120\,R_{\mu\nu\rho\sigma}R^{\mu\nu\rho}_{\,\,\,\,\,\,\,\,\,\alpha}R^{\sigma\alpha}
\right.\nonumber\\&&\left.
144\,R_{\mu\nu\rho\sigma}R^{\mu\rho}R^{\nu\sigma}+128
R_{\mu}^{\,\,\,\,\nu}R_{\nu}^{\,\,\,\,\rho}R_{\rho}^{\,\,\,\,\mu}-108
R_{\mu}^{\,\,\,\,\nu}R_{\nu}^{\,\,\,\,\mu}R+11\,R^3\right).
\end{eqnarray}%
The planar AdS black hole solutions for different values of the
coupling constants were found in \cite{Myers:2010ru}. The
solution, in units where the radius of $AdS$ is one, is %
\be %
ds^2=r^2\left(-N^2\,f(r)dt^2+d\vec{x}^2\right)+\frac{dr^2}{r^2\,f(r)},\label{metric}
\ee%
where $f(r)$ is determined by roots of the
following equation%
\be 1-f(r)+\lambda f(r)^2+\mu f(r)^3=\frac{r_h^4}{r^4}. \label{fr}\ee%
Here $r$ denotes the radial coordinate of the black brane geometry
and $t, \vec{x}$ label the directions along the boundary at the
spatial infinity. In these coordinates the event horizon is located
at $f(r_h)=0$ where $r_h$ is found by solving this equation. The
boundary is located at infinity and the geometry will be as
asymptotically AdS . The constant $N^2$ specifies the speed of light
of the boundary gauge theory and one can choose it to be unity. We
name $f(r)$ at the boundary where $r\rightarrow\infty$, as $f_{\infty}$ and one finds that%
\be N^2=\frac{1}{f_\infty},\ee%
One also finds from \eqref{fr} that $f_{\infty}$ satisfies%
\be 1-f_{\infty}+\lambda f_{\infty}^2+\mu f_{\infty}^3=0.\ee%
The temperature of the hot plasma is given by the Hawking
temperature of the black hole
\begin{equation}
T=\frac{N\,r_h}{\pi}\label{T}.
\end{equation}
Authors in \cite{Myers:2010ru}, solved \eqref{fr} and found $f(r)$
for different values of coupling constants $\lambda$ and $\mu $. It
is shown that there are three different solutions of \eqref{fr} in
the $\mu-\lambda$ plane:%
\bea f_1(r)&=&u+v-\frac{\lambda}{3\mu},\nonumber\\
f_2(r)&=&-\frac{u+v}{2}+i\,\frac{\sqrt{3}}{2}(u-v)-\frac{\lambda}{3\mu},
\nonumber \\
f_3(r)&=&-\frac{u+v}{2}-i\,\frac{\sqrt{3}}{2}(u-v)-\frac{\lambda}{3\mu},\label{f}\eea%
where
\be
u=(q+\sqrt{q^2-p^3})^{\frac{1}{3}},\,\,\,\,\,\,v=(q-\sqrt{q^2-p^3})^{\frac{1}{3}},
\ee
and
\be
p=\frac{3\mu+\lambda^2}{9\mu^2},\,\,\,q=-\frac{2\lambda^3+9\mu\lambda+27\mu^2
\left(1-\frac{r_h^4}{r^4}\right)}{54\mu^3}.\label{p} \ee
There is a relation between Gauss-Bonnet coupling constant $\lambda$
and cubic-curvature coupling constant $\mu$ as follows%
\be \mu=\frac{2}{27}-\frac{\lambda}{3}\pm\frac{2}{27}\left(1-3\lambda\right)^{\frac{3}{2}}.\ee%
which shows the upper and lower bound on the cubic-curvature
interaction coupling. There is a special case $p=0$ in \eqref{p}
which corresponds to $\mu=-\frac{\lambda^2}{3}$. $f(r)$ is also
found at this point.

%%%%%%%%%%%%%%%
%%%%%%%%%%%%%%%%%%%
%%%%%%%%%%%%%%%%%%%
%%%%%%%%%%%%

\end{document}